\documentclass[12pt, draftcls, onecolumn,journal]{IEEEtran}

\makeatletter
\let\NAT@parse\undefined
\makeatother

\makeatletter
\def\markboth#1#2{\def\leftmark{\@IEEEcompsoconly{\sffamily}\MakeUppercase{\protect#1}}%
\def\rightmark{\@IEEEcompsoconly{\sffamily}\MakeUppercase{\protect#2}}}
\makeatother

\usepackage{epsfig,makeidx,color}
\usepackage{amsmath,mathrsfs}
\usepackage{amssymb,amsfonts}
\usepackage{amsthm} 
\usepackage[english]{babel}
\usepackage[latin1]{inputenc}
\usepackage{graphicx,color}
\usepackage{amscd}
\usepackage{verbatim}
\usepackage{float}
\usepackage{dsfont}
\usepackage{citesort}
\usepackage[square, comma, numbers]{natbib}
\usepackage[ruled]{algorithm2e}

\DeclareMathOperator*{\argmin}{argmin}

\providecommand{\abs}[1]{\ensuremath{\left\lvert #1 \right\rvert}}
\providecommand{\norm}[1]{\ensuremath{\left\Vert #1 \right\Vert}}
\providecommand{\dotproduct}[2]{\ensuremath{\left< #1, #2 \right>}}

\providecommand{\nint}[1]{\ensuremath{\left\lfloor #1 \right \rceil}}

\providecommand{\vv}[1]{\textquotedblleft #1\textquotedblright}

\newcommand{\Z}{\mathbb{Z}}
\newcommand{\C}{\mathbb{C}}
\newcommand{\R}{\mathbb{R}}

\DeclareMathOperator*{\Span}{span}

\DeclareMathOperator*{\dB}{dB}
\DeclareMathOperator*{\ML}{ML}
\DeclareMathOperator*{\red}{red}
\DeclareMathOperator*{\ALR}{ALR}
\DeclareMathOperator*{\SNR}{SNR}
\DeclareMathOperator*{\cc}{c}
\providecommand{\abs}[1]{\ensuremath{\left\lvert #1 \right\rvert}}
\providecommand{\norm}[1]{\ensuremath{\left\Vert #1 \right\Vert}}

\providecommand{\vv}[1]{\textquotedblleft #1\textquotedblright}

\newtheorem{lem}{Lemma}
\newtheorem{prop}{Proposition}

\newtheorem*{defn}{Definition}
\theoremstyle{definition}

\newtheorem*{rem}{Remark}

\begin{document}
\title{Augmented Lattice Reduction \\ for MIMO decoding}
\author{L. Luzzi $\qquad$ G. Rekaya-Ben Othman $\qquad$ J.-C. Belfiore%
\thanks{Jean-Claude Belfiore, Ghaya Rekaya-Ben Othman and Laura Luzzi are with Télécom-ParisTech, 46 Rue\ %
Barrault, 75013 Paris, France. E-mail: $\tt \{belfiore,rekaya,luzzi\}$@$\tt telecom-paristech.fr$. Tel: +33 (0)145817705, +33 (0)145817633, +33 (0)145817636. Fax: +33 (0)145804036 \par
\copyright \textbf{ This work has been submitted to the IEEE for possible publication. Copyright may be transferred without notice, after which this version may no longer be accessible.}
}}
\date{January 5, 2010}
\bibliographystyle{IEEE}

\maketitle

\begin{abstract}
Lattice reduction algorithms, such as the LLL algorithm, have been proposed as preprocessing tools in order to enhance the performance of suboptimal receivers in MIMO communications.\\
In this paper we introduce a new kind of lattice reduction-aided decoding technique, called \emph{augmented lattice reduction}, which recovers the transmitted vector directly from the change of basis matrix, and therefore doesn't entail the computation of the pseudo-inverse of the channel matrix or its QR decomposition.\\
We prove that augmented lattice reduction attains the maximum receive diversity order of the channel; simulation results evidence that it significantly outperforms LLL-SIC detection without entailing any additional complexity. A theoretical bound on the complexity is also derived. 
\medskip \par
\textbf{Index Terms}: lattice reduction-aided decoding, LLL algorithm, right preprocessing. 
\end{abstract}

\section{Introduction}
Multiple-input multiple-output (MIMO) systems can
provide high data rates and reliability over fading channels. In order to achieve optimal performance, maximum likelihood decoders such as the Sphere Decoder may be employed; however, their complexity grows prohibitively with the number of antennas and the constellation size, posing a challenge for practical implementation. \\
On the other hand, suboptimal receivers such as zero forcing (ZF) or successive interference cancellation (SIC) do not preserve the diversity order of the system
\cite{KCM}.
Right preprocessing using \emph{lattice reduction} 
has been proposed 
in order to enhance their performance 
\cite{YW, DEC2, WF}. In particular, the classical LLL algorithm for lattice reduction,
whose average complexity is polynomial in the number of antennas\footnote{Note that the \emph{worst-case} number of iterations of the LLL algorithm applied to the MIMO context is unbounded, as has been proved in \cite{JSM}. However, the tail probability of the number of iterations decays exponentially, so that in many cases high complexity events can be regarded as negligible with respect to the target error rate (see \cite{JE}, Theorem 3).}, has been proven to achieve the optimal receive diversity order in the spatial multiplexing  
case \cite{TMK}. Very recently, it has also been shown that combined with regularization techniques such as MMSE-GDFE left preprocessing, lattice reduction-aided decoding is optimal in terms of diversity-multiplexing tradeoff \cite{JE}. However, the shift between the error probability of ML detection and LLL-ZF (respectively, LLL-SIC) detection increases greatly for a large number of antennas \cite{Li}.\\
In this paper we present a new kind of LLL-aided decoding, called \emph{augmented lattice reduction}, which doesn't require ZF or SIC receivers and therefore doesn't entail the computation of the pseudo-inverse of the channel matrix or its QR decomposition.\\
In the coherent case, MIMO decoding amounts to solving an instance of the \emph{closest vector problem} (CVP) in a finite subset of the lattice generated by the channel matrix\footnote{Actually,  LLL-ZF and LLL-SIC suboptimal decoding correspond to two classical techniques for finding approximate solutions of the CVP, due to Babai: the \emph{rounding algorithm} and \emph{nearest plane algorithm} respectively \cite{Ba}.}. Following an idea of Kannan \cite{Ka}, our strategy is to reduce the CVP to the \emph{shortest vector problem} (SVP) by embedding the $n$-dimensional lattice generated by the channel matrix into an $(n+1)$-dimensional lattice. We show that for a suitable choice of the embedding, the transmitted message can be recovered directly from the coordinates of the shortest vector of the augmented lattice.\\
In general, the LLL algorithm is not guaranteed to solve the SVP; however, it certainly finds the shortest vector in the lattice in the particular case where the minimum distance is exponentially smaller than the other successive minima. Equivalently, we can say that \vv{the LLL algorithm is an \emph{SVP-oracle} when the lattice gap is exponential in the lattice dimension}. An appropriate choice of the embedding ensures that this condition is satisfied. \\
Thanks to this property, we can prove that our method also achieves the receive diversity of the channel. Numerical simulations evidence that augmented lattice reduction significantly outperforms LLL-SIC detection without entailing any additional complexity. A theoretical (albeit pessimistic) bound on the complexity is also derived.
\medskip \par
This paper is organized as follows: in Section \ref{preliminaries} we introduce the system model and basic notions concerning lattice reduction, and summarize the existing lattice reduction-aided decoding schemes. In Section \ref{algorithm} we describe augmented lattice reduction decoding, and in Section \ref{performance} we analyze its performance and complexity, both theoretically and through numerical simulations.  

\section{Preliminaries} \label{preliminaries}

\subsection{System model and notation} \label{system_model}
We consider a MIMO system with $M$ transmit and $N$ receive antennas such that $M\leq N$ using spatial multiplexing. The complex received signal is given by
\begin{equation} \label{channel}
\mathbf{y}_{\cc}=\mathbf{H}_{\cc}\mathbf{x}_{\cc}+\mathbf{w}_{\cc}, 
\end{equation}
where $\mathbf{x}_{\cc}\in\C^M$, $\mathbf{y}_{\cc}$, $\mathbf{w}_{\cc}\in\C^N$, $\mathbf{H}_{\cc}\in M_{N\times M}(\C)$. 
The transmitted vector $\mathbf{x}_{\cc}$ belongs to a finite constellation $\mathcal{S} \subset \Z[i]^M$; the entries of the channel matrix $\mathbf{H}_{\cc}$ are supposed to be i.i.d. complex Gaussian random variables with zero mean and variance per real dimension equal to $\frac{1}{2}$, and $\mathbf{w}_{\cc}$ is the Gaussian noise with i.i.d. entries of zero mean and variance $N_0$. We consider the coherent case where $\mathbf{H}_{\cc}$ is known at the receiver.\\
Separating the real and imaginary part, the model can be rewritten as 
\begin{equation} \label{real_system}
\mathbf{y}=\mathbf{H}\mathbf{x}+\mathbf{w},
\end{equation}
in terms of the real-valued vectors 
$$\mathbf{y}=\begin{pmatrix} \Re(\mathbf{y}_{\cc}) \\ \Im(\mathbf{y}_{\cc}) \end{pmatrix} \in \R^n, \quad \mathbf{x}=\begin{pmatrix} \Re(\mathbf{x}_{\cc}) \\ \Im(\mathbf{x}_{\cc}) \end{pmatrix} \in \Z^m$$
and of the equivalent real channel matrix 
$$\mathbf{H}=\begin{pmatrix} \Re(\mathbf{H}_{\cc}) & -\Im(\mathbf{H}_{\cc}) \\ 
                                                          \Im(\mathbf{H}_{\cc}) & \Re(\mathbf{H}_{\cc}) \end{pmatrix} \in M_{n \times m} (\R).$$
Here $n=2N$, $m=2M$.\\
The maximum likelihood decoded vector is given by
\begin{equation*}
\hat{\mathbf{x}}_{\ML}=\argmin_{\hat{\mathbf{x}}_{\cc} \in \mathcal{S}} \norm{\mathbf{H}_{\cc}\hat{\mathbf{x}}_{\cc}-\mathbf{y}_{\cc}}= \argmin_{\hat{\mathbf{x}}_{\cc} \in \mathcal{S}} \norm{\mathbf{H}\hat{\mathbf{x}}-\mathbf{y}},
\end{equation*} 
where $\norm{\cdot}$ denotes the Euclidean norm. 

\subsection{Lattice reduction} \label{LR}
An $m$-dimensional real 
lattice in $\R^n$ is the set of points 
$$\mathcal{L}(\mathbf{H})=\{\mathbf{Hx} \; | \; \mathbf{x} \in \Z^m\},$$
where $\mathbf{H} \in  M_{n\times m}(\R)$. We denote by $d_{\mathbf{H}}$ the \emph{minimum distance} of the lattice, that is the smallest norm  of a nonzero vector in $\mathcal{L}(\mathbf{H})$. More generally, for all $1 \leq i \leq m$ one can define the \emph{$i$-th successive minimum} of the lattice as follows:
\begin{multline*}
\lambda_i(\mathbf{H})=\inf\{r>0 \;|\; \exists \mathbf{v}_1,\ldots,\mathbf{v}_i \text{ linearly independent } \text{in } \mathcal{L}(\mathbf{H}) \text{ s.t. }
\norm{\mathbf{v}_j} \leq r \quad \forall j \leq i\}
\end{multline*} 
We recall that two matrices $\mathbf{H}, \mathbf{H'}$ generate the same lattice if and only if $\mathbf{H}'=\mathbf{H}\mathbf{U}$ with $\mathbf{U}$ unimodular.\\
\emph{Lattice reduction} algorithms allow to find a new basis $\mathbf{H}'$ for a given lattice $\mathcal{L}(\mathbf{H})$ such that the basis vectors are shorter and nearly orthogonal. Orthogonality can be measured by the absolute value of the coefficients $\mu_{i,j}$ in the Gram-Schmidt orthogonalization of the basis, see the GSO Algorithm \ref{GSO}. 

\begin{algorithm} \label{GSO}
\DontPrintSemicolon
\SetAlgoLined
$\mathbf{h}_1^* \leftarrow \mathbf{h}_1$\;
\For{$i=2,\ldots,m$}{
\For{$j=1,\ldots,i-1$}
{$\mu_{i,j} \leftarrow \frac{\dotproduct{\mathbf{h}_i}{\mathbf{h}_j^*}}{\norm{\mathbf{h}_j^*}^2}$\;}
$\mathbf{h}_i^* \leftarrow \mathbf{h}_i-\sum_{j=1}^{i-1} \mu_{i,j} \mathbf{h}_j^*$}
\caption{GSO (Gram-Schmidt orthogonalization)}
\end{algorithm}

We recall the following useful property of GSO: the length of the smallest of the Gram-Schmidt vectors $\mathbf{h}_i^*$ is always less or equal to the minimum distance $d_{\mathbf{H}}$ of the lattice \cite{LLS}. In other words,
\begin{equation} \label{lambda}
d_{\mathbf{H}} \geq a(\mathbf{H})\doteqdot\min_{1\leq i \leq m} \norm{\mathbf{h}_i^*}
\end{equation}

A basis $\mathbf{H}$ is said to be \emph{LLL-reduced} \cite{LLL} if its Gram-Schmidt coefficients $\mu_{i,j}$ and Gram-Schmidt vectors  satisfy the following properties:
\begin{enumerate}
\item \emph{Size reduction:} $$\abs{\mu_{k,l}}<\frac{1}{2}, \quad 1 \leq l < k \leq m,$$
\item \emph{Lovasz condition:} \quad 
$$\norm{\mathbf{h}_k^*+\mu_{k,k-1} \mathbf{h}_{k-1}^*}^2 \geq \delta \norm{\mathbf{h}_{k-1}^*}^2, \quad 1 < k \leq m,$$
\end{enumerate}
where $\delta \in \left(\frac{1}{4},1\right)$ (a customary choice is $\delta=\frac{3}{4}$).\\
The LLL algorithm is summarized in Algorithm \ref{LLL_algorithm}. Given a full-rank matrix $\mathbf{H} \in M_{n \times m} (\R)$, it computes an LLL-reduced version $\mathbf{H}_{\red}=\mathbf{HU}$, with $\mathbf{U} \in M_{m \times m}(\Z)$ unimodular, and outputs the columns $\{\mathbf{h}_i\}$ and $\{\mathbf{u}_i\}$ of $\mathbf{H}_{\red}$ and $\mathbf{U}$ respectively.
\begin{algorithm} \label{LLL_algorithm}
\DontPrintSemicolon
\SetAlgoLined
$\mathbf{U}=\mathbf{I}_{m}$\;
Compute the GSO of $\mathbf{H}$\;
$k \leftarrow 2$\;
\While{$k\leq m$}{
RED(k,k-1)\;
\If{$\norm{\mathbf{h}_k^*+\mu_{k,k-1}\mathbf{h}_{k-1}^*}^2 < \delta \norm{\mathbf{h}_{k-1}^*}^2$}
{
swap $\mathbf{h}_k$ and $\mathbf{h}_{k-1}$\;
swap $\mathbf{u}_k$ and $\mathbf{u}_{k-1}$\;
update GSO\;
$k \leftarrow \max(k-1,2)$\;
}
\Else
{
\For{$l=k-2,\ldots,1$}
{
RED(k,l)
}
$k \leftarrow k+1$\;
}
}
\caption{The LLL algorithm}
\end{algorithm}

\begin{algorithm}
\DontPrintSemicolon
\SetAlgoLined
\If{$\abs{\mu_{k,l}}>\frac{1}{2}$}
{
$\mathbf{h}_k \leftarrow \mathbf{h}_k -\nint{\mu_{k,l}}\mathbf{h}_l$\;
$\mathbf{u}_k \leftarrow \mathbf{u}_k -\nint{\mu_{k,l}}\mathbf{u}_l$\;
\For{$j=1,\cdots,l-1$}{
$\mu_{k,j} \leftarrow \mu_{k,j}-\nint{\mu_{k,l}}\mu_{l,j}$
} 
$\mu_{k,l} \leftarrow \mu_{k,l}-\nint{\mu_{k,l}}$
}
\caption{Size reduction RED(k,l)}
\end{algorithm}

We list here some properties of LLL-reduced bases that we will need in the sequel. First of all, the LLL algorithm finds at least one basis vector whose length is not too far from the minimum distance $d_{\mathbf{H}}$ of the lattice. The following inequality holds for any $m$-dimensional LLL-reduced basis $\mathbf{H}$ \cite{Co}:
\begin{equation} \label{first_column}
\norm{\mathbf{h}_1} \leq \alpha^{\frac{m-1}{2}} d_{\mathbf{H}},
\end{equation}
where $\alpha=\frac{1}{\delta-1/4}$ ($\alpha=2$ if $\delta=\frac{3}{4}$). \\
Moreover, the first basis vector cannot be too big compared to the Gram-Schmidt vectors $\{\mathbf{h}_i^*\}$:
$$ \norm{\mathbf{h}_1} \leq \alpha^{\frac{i-1}{2}} \norm{\mathbf{h}_i^*}, \quad \quad \forall 1 \leq i \leq m.$$
In particular, if $j=\argmin_{1 \leq i \leq m} \norm{\mathbf{h}_i^*}$,
\begin{equation} \label{B}
d_{\mathbf{H}} \leq \norm{\mathbf{h}_1} \leq \alpha^{\frac{j-1}{2}} \norm{\mathbf{h}_j^*}= \alpha^{\frac{j-1}{2}} a(\mathbf{H}) \leq \alpha^{\frac{m-1}{2}} a(\mathbf{H}).
\end{equation}

\subsection{Lattice reduction-aided decoding}
In this section we briefly review existing detection schemes which use the LLL algorithm to preprocess the channel matrix, in order to improve the performance of suboptimal decoders such as ZF or SIC \cite{YW, WF, DEC2}.\\
Let $\mathbf{H}_{\red}=\mathbf{HU}$ be the output of the LLL algorithm on $\mathbf{H}$. We can rewrite the received vector as  $\mathbf{y}=\mathbf{H}_{\red}\mathbf{U}^{-1}\mathbf{x}+\mathbf{w}$. 
\begin{itemize}
\item The \emph{LLL-ZF decoder} outputs 
$$\mathbf{\hat{x}}_{LLL-ZF}=Q_{\mathcal{S}}\left(\mathbf{U}\left(\nint{\mathbf{H}_{\red}^{\dagger}\mathbf{y}}\right)\right),$$
where $\mathbf{H}_{\red}^{\dagger}=(\mathbf{H}_{\red}^T \mathbf{H}_{\red})^{-1}\mathbf{H}_{\red}^T$ is the Moore-Penrose pseudoinverse of $\mathbf{H}_{\red}$, $\nint{\cdot}$ denotes componentwise rounding to the nearest integer and $Q_{\mathcal{S}}$ is a quantization function that forces the solution to belong to the constellation $\mathcal{S}$.
\item The \emph{LLL-SIC decoder} performs the QR decomposition $\mathbf{H}_{\red}=\mathbf{Q}\mathbf{R}$, computes $\widetilde{\mathbf{y}}=\mathbf{Q}^T\mathbf{y}$, finds by recursion $\widetilde{\mathbf{x}}$ defined by
\begin{align*}
&\tilde{x}_m=\nint{\frac{\tilde{y}_m}{r_{mm}}},\\
& \tilde{x}_i=\nint{\frac{\tilde{y}_i-\sum_{j=i+1}^{m} r_{ij}\tilde{x}_j}{r_{ii}}}, \qquad i=m-1,\ldots,1,
\end{align*} 
and finally outputs $\mathbf{\hat{x}}_{LLL-SIC}=Q_{\mathcal{S}}\left(\mathbf{U}\widetilde{\mathbf{x}}\right)$.
\end{itemize}

\section{Augmented Lattice Reduction} \label{algorithm}
We propose here a new decoding technique based on the LLL algorithm which, unlike the LLL-ZF and LLL-SIC decoders, does not require the inversion of the channel matrix at the last stage. Let $\mathbf{y}$ be the (real) received vector in the model (\ref{real_system}). 
Consider the $(n+1)\times(m+1)$ augmented matrix
\begin{equation} \label{Htilde}
\widetilde{\mathbf{H}}=\begin{pmatrix} \mathbf{H} & -\mathbf{y} \\ \mathbf{0}_{1\times m} & t \end{pmatrix}= \left(\begin{array}{cccc} h_{1,1} & \cdots & h_{1,m} & -y_1 \\ 
                                                 \vdots & & & \vdots \\
                                                 h_{n,1} & \cdots & h_{n,m} & -y_n \\
                                                   0    & \cdots        & 0   & t  \end{array} \right)
\end{equation}
where $t>0$ is a parameter to be determined. The points of the augmented lattice $\mathcal{L}(\widetilde{\mathbf{H}})$ are of the form 
$$\begin{pmatrix} \mathbf{H}\mathbf{x'}-q\mathbf{y} \\ qt \end{pmatrix}, \quad \quad \mathbf{x}'\in \Z^m, \; q \in \Z$$
In particular, the vector
$\mathbf{v}=\begin{pmatrix} \mathbf{H}\mathbf{x}-\mathbf{y} \\ t \end{pmatrix}=\begin{pmatrix} \mathbf{w} \\ t \end{pmatrix}$
belongs to the augmented lattice. We will show that for a suitable choice of the parameter $t$, and supposing that the noise is small enough, $\mathbf{v}$ is the shortest vector in the lattice and the LLL algorithm finds this vector. That is, $\pm \mathbf{v}$ is the first column of $\widetilde{\mathbf{H}}_{\red}=\widetilde{\mathbf{H}}\widetilde{\mathbf{U}}$, the output of LLL algorithm on $\widetilde{\mathbf{H}}$. Clearly, since $\widetilde{\mathbf{H}}$ is full-rank with probability $1$, in this case the first column of the change of basis matrix $\widetilde{\mathbf{U}}$ is $\begin{pmatrix} \pm \mathbf{x} \\ \pm 1 \end{pmatrix}$. Thus we can \vv{read} the transmitted message directly from the change of basis matrix $\widetilde{\mathbf{U}}$.\\
To summarize, in order to decode we can perform the LLL algorithm on $\widetilde{\mathbf{H}}$, and given the output $\widetilde{\mathbf{H}}_{\red}=\widetilde{\mathbf{H}}\widetilde{\mathbf{U}}$, we can choose
\begin{equation} \label{xLag1}
\hat{\mathbf{x}}=Q_{\mathcal{S}}\left(\nint{\frac{1}{\widetilde{u}_{m+1,1}}(\widetilde{u}_{1,1},\ldots,\widetilde{u}_{m,1})^T}\right),
\end{equation}
where $\widetilde{\mathbf{U}}=(\widetilde{u}_{i,j})$.\\
The previous decoder can be improved by including all the columns of $\mathbf{H}_{\red}$ in the search for the vector $\mathbf{v}$. Specifically, let
$$\mathbf{u}_k=\frac{1}{\widetilde{u}_{m+1,k}}(\widetilde{u}_{1,k},\ldots,\widetilde{u}_{m,k})^T, \quad k=1,\ldots,m.$$
If there exists some $k \in \{1,\ldots,m\}$ such that \abs{\widetilde{u}_{m+1,k}}=1, we define
$$k_{\min}=\argmin_{k  \text{ s.t. } \abs{\widetilde{u}_{m+1,k}}=1} \norm{\mathbf{H}\mathbf{u}_k-\mathbf{y}},$$
otherwise $k_{\min}=1$. Then the \emph{Augmented Lattice Reduction decoder} outputs
\begin{equation} \label{xLag2}
\hat{\mathbf{x}}_{\ALR}=Q_{\mathcal{S}}\left(\nint{\mathbf{u}_{k_{\min}}}\right),
\end{equation}

\section{Performance} \label{performance}
\subsection{Diversity} \label{diversity}
In this paragraph we will investigate the performance of augmented lattice reduction. We begin by proving that our method, like LLL-ZF and LLL-SIC, attains the maximum receive diversity gain of $N$, for an appropriate choice of the parameter $t$ in (\ref{Htilde}). The diversity gain $d$ of a decoding scheme is defined as follows:  
$$d=-\lim_{\rho \to \infty} \frac{\log(P_e)}{\log(\rho)},$$
where $P_e$ denotes the error probability as a function of the signal to noise ratio $\rho$.

\begin{prop} \label{receive_diversity}
If the augmented lattice reduction is performed using $t=\varepsilon a(\mathbf{H}_{\red})$, where $a(\mathbf{H}_{\red})$ is the length of the smallest vector in the Gram-Schmidt orthogonalization of $\mathbf{H}_{\red}$, and $\varepsilon\leq\frac{1}{2\sqrt{2} \alpha^{m-\frac{1}{2}}}$, then it achieves the maximum receive diversity $N$. 
\end{prop}

\begin{rem}
It is essential to use $a(\mathbf{H}_{\red})$ in place of $a(\mathbf{H})$. In fact, for general bases $\mathbf{H}$ that are not LLL-reduced, there is no lower bound of the type (\ref{B}) limiting how small the smallest Gram-Schmidt vector can be. For $a(\mathbf{H}_{\red})$, putting together the bounds (\ref{lambda}) and (\ref{B}), we obtain
\begin{equation} \label{AB}
\frac{d_{\mathbf{H}}}{\alpha^{\frac{m-1}{2}}} \leq a(\mathbf{H}_{\red}) \leq d_{\mathbf{H}}
\end{equation}
Note that the LLL reduction of $\mathbf{H}$ does not entail any additional complexity, since it is the same as the LLL reduction on the first $m$ columns of $\widetilde{\mathbf{H}}$. In fact the parameter $t$ can be chosen during the LLL reduction of $\widetilde{\mathbf{H}}$, after carrying out the LLL algorithm on the first $m$ columns.
\end{rem}


In order to prove the previous Proposition, we will show that in the $(m+1)$-dimensional lattice $\mathcal{L}(\widetilde{\mathbf{H}})$ there is an exponential gap between the first two successive minima. Then, using the estimate (\ref{first_column}) on the norm of the first vector in an LLL-reduced basis, one can conclude that in this particular case the LLL algorithm finds the shortest vector in the lattice $\mathcal{L}(\widetilde{\mathbf{H}})$ with high probability. This, in turn, allows to recover the closest lattice vector $\mathbf{Hx}$ to $\mathbf{y}$ in $\mathcal{L}(\mathbf{H})$ supposing that the noise $\mathbf{w}$ is small enough.\\
The following definition makes the notion of \vv{gap} more precise:

\begin{defn}
Let $\mathbf{v}$ be a shortest nonzero vector in the lattice $\mathcal{L}(\mathbf{H})$, and let $\gamma>1$. $\mathbf{v}$ is called \emph{$\gamma$-unique} if $\forall \mathbf{u} \in \mathcal{L}(\mathbf{H})$,
$$ \norm{\mathbf{u}} \leq \gamma \norm{\mathbf{v}} \quad \Rightarrow \quad \mathbf{u}, \mathbf{v} \quad \text{are linearly dependent.}$$
\end{defn}

We now prove the existence of such a gap under suitable conditions:

\begin{lem} \label{Lemma32}
Let $\widetilde{\mathbf{H}}$ be the matrix defined in (\ref{Htilde}), and let $t=\varepsilon a(\mathbf{H}_{\red})$, with $\varepsilon \leq \frac{1}{2\sqrt{2}\alpha^{m-\frac{1}{2}}}$.\\ 
Suppose that
$\norm{\mathbf{w}}=\norm{\mathbf{y}-\mathbf{H}\mathbf{x}}\leq \varepsilon d_{\mathbf{H}}$.\\
Then $\mathbf{v}=\begin{pmatrix} \mathbf{Hx-y} \\ t\end{pmatrix}$ is an $\alpha^{\frac{m}{2}}$-unique shortest vector of $\mathcal{L}(\widetilde{\mathbf{H}})$. 
\end{lem}

\begin{rem} Observe that the hypothesis on $\norm{\mathbf{w}}$ implies in particular that $\norm{\mathbf{w}}<\frac{d_{\mathbf{H}}}{2}$ and $\mathbf{Hx}$ is indeed the closest lattice point to $\mathbf{y}$.
\end{rem}

\begin{IEEEproof}
We need to show that any vector $\mathbf{u} \in \mathcal{L}(\widetilde{\mathbf{H}})$ that is not a multiple of $\mathbf{v}$ must have length greater than $\alpha^{\frac{m}{2}}\norm{\mathbf{v}}$.\\
By contradiction, suppose that $\exists \mathbf{u}=\begin{pmatrix}\mathbf{Hx'}-q\mathbf{y} \\ qt\end{pmatrix} \in \mathcal{L}(\widetilde{\mathbf{H}})$ linearly independent from $\mathbf{v}$ such that $\norm{\mathbf{u}} \leq \alpha^{\frac{m}{2}}\norm{\mathbf{v}}.$ 
Since $\norm{\mathbf{u}} \geq \abs{q} t$, 
$$\abs{q} \leq \frac{\norm{\mathbf{u}}}{t} \leq \frac{\alpha^{\frac{m}{2}}\norm{\mathbf{v}}}{t}.$$
On the other side, $\norm{\mathbf{u}} \leq \alpha^{\frac{m}{2}}\norm{\mathbf{v}}$ implies that also $\norm{\mathbf{Hx'}-q\mathbf{y}} \leq \alpha^{\frac{m}{2}}\norm{\mathbf{v}}$. Consider
\begin{align}
&\norm{\mathbf{Hx'}-q\mathbf{Hx}} = \norm{\mathbf{Hx'}-q\mathbf{y}} + \norm{q\mathbf{y}-q\mathbf{Hx}} \leq \notag\\
&\leq \alpha^{\frac{m}{2}}\norm{\mathbf{v}} + \abs{q} \norm{\mathbf{y}-\mathbf{Hx}} \leq \alpha^{\frac{m}{2}}\norm{\mathbf{v}} + \frac{\alpha^{\frac{m}{2}}\norm{\mathbf{v}}}{t} \norm{\mathbf{w}} \leq \notag \\
&\leq \alpha^{\frac{m}{2}} \sqrt{\norm{\mathbf{w}}^2+t^2}\left(1+\frac{\norm{\mathbf{w}}}{t}\right) \label{ineq}
\end{align}
The bound (\ref{AB}) on $a(\mathbf{H}_{\red})$ implies
$$\frac{\varepsilon}{\alpha^{\frac{m-1}{2}}} d_{\mathbf{H}} \leq t \leq \varepsilon d_{\mathbf{H}}.$$
Using this inequality and the hypotheses on $\norm{\mathbf{w}}$ and $\varepsilon$, we can bound the expression (\ref{ineq}) with
$$\alpha^\frac{m}{2}\sqrt{2}\varepsilon d_{\mathbf{H}} \left(1+ \alpha^{\frac{m-1}{2}}\right) < 2\sqrt{2} \varepsilon d_{\mathbf{H}} \alpha^{\frac{m}{2}}\alpha^{\frac{m-1}{2}}\leq d_{\mathbf{H}}.$$
Thus $\norm{\mathbf{Hx'}-q\mathbf{Hx}}<d_{\mathbf{H}}$. But this is a contradiction because $\mathbf{Hx'}-q\mathbf{Hx} \in \mathcal{L}(\mathbf{H})$ and is nonzero since $\mathbf{v}$ and $\mathbf{u}$ are linearly independent. Therefore $\mathbf{v}$ is $\alpha^{\frac{m}{2}}$-unique. (Since the last coordinate of $\mathbf{v}$ in the basis $\widetilde{\mathbf{H}}$ is $1$, $\mathbf{v}$ cannot be a nontrivial multiple of another lattice vector.) 
\end{IEEEproof}

\begin{rem}
The lower bound on $t$ is essential to ensure that $\abs{q}$ is bounded. If $\abs{q}$ were unbounded, clearly $\norm{\mathbf{H}\mathbf{x}'-q\mathbf{y}}$ might be arbitrarily small and there might exist $\mathbf{u} \in \mathcal{L}(\widetilde{\mathbf{H}})$ of smaller norm than $\mathbf{v}$.
\end{rem}

\begin{lem}
Under the hypotheses of Lemma \ref{Lemma32}, the augmented lattice reduction methods (\ref{xLag1}) and (\ref{xLag2}) correctly decode the transmitted signal $\mathbf{x}$.
\end{lem}

\begin{IEEEproof}
Let $\widetilde{\mathbf{H}}_{\red}=\widetilde{\mathbf{H}}\widetilde{\mathbf{U}}$ denote the output of the LLL reduction of $\widetilde{\mathbf{H}}$, and let $\hat{\mathbf{h}}_1=\widetilde{\mathbf{H}}\begin{pmatrix} \mathbf{x}' \\ q \end{pmatrix}$ be its first column. The property (\ref{first_column}) of LLL reduction in dimension $m+1$ entails that $\norm{\hat{\mathbf{h}}_1} \leq {\alpha}^{\frac{m}{2}}d_{\widetilde{\mathbf{H}}}$. But since $\mathbf{v}=\begin{pmatrix} \mathbf{Hx-y} \\ t \end{pmatrix}$ has been shown to be ${\alpha}^{\frac{m}{2}}$-unique in the previous Lemma, it means that $\hat{\mathbf{h}}_1$ and $\mathbf{v}$ are linearly dependent; equivalently, $\exists a,b \in \Z \setminus \{0\}$ such that $a\mathbf{v}+b\hat{\mathbf{h}}_1=0$. In particular $a t +b q t =0$, that is $a=-b q$ and $\hat{\mathbf{h}}_1=q\mathbf{v}$. Then by definition of $\widetilde{\mathbf{H}}$, $$\hat{\mathbf{h}}_1=\widetilde{\mathbf{H}}\begin{pmatrix} q\mathbf{x}\\ q \end{pmatrix}.$$
 This means that the first column  of the reduction matrix $\widetilde{\mathbf{U}}$ is $\begin{pmatrix} q\mathbf{x} \\ q\end{pmatrix}$, and so $\hat{\mathbf{x}}_{\ALR}=Q_{\mathcal{S}}(\nint{\mathbf{u}_1})= Q_{\mathcal{S}}\left(q\mathbf{x}/q\right)=\mathbf{x}$ and the augmented lattice reduction methods (\ref{xLag1}) and (\ref{xLag2}) correctly decode the transmitted message.\\
(Observe that this is possible only if $\abs{q}=1$, since $\det(\widetilde{\mathbf{U}})$ is also a multiple of $q$ and $\widetilde{\mathbf{U}}$ is unimodular.)
\end{IEEEproof} 
\medskip \par
Thus for any channel realization $\mathbf{H}$, we have the following bound on the error probability for the augmented lattice reduction method:
$$P_{e,\ALR}(\mathbf{H}) \leq P\{\norm{\mathbf{w}}>\varepsilon d_{\mathbf{H}}\}.$$
To conclude the proof of Proposition \ref{receive_diversity}, we need to show that given $\varepsilon \leq \frac{1}{2\sqrt{2}\alpha^{m-\frac{1}{2}}}$, we have 
$$\lim_{\rho \to \infty} \frac{-\log P\{\norm{\mathbf{w}}> \varepsilon d_{\mathbf{H}}\}}{\log \rho} \geq N$$
This turns out to be true. In fact, it has been shown in \cite{TMK} (Proof of Theorem 2), that for any constant $c_M$ depending only on the number of transmit antennas\footnote{This result was used in \cite{TMK} in order to prove that the LLL-ZF decoder achieves the receive diversity order. The proof in \cite{TMK} actually refers to the complex model (\ref{channel}), but the statement also holds for the real model since $d_{\mathbf{H}}=d_{\mathbf{H}_{\cc}}$, $\norm{\mathbf{w}}=\norm{\mathbf{w}_{\cc}}$.},
\begin{align*}
&P\{\norm{\mathbf{w}}>c_M d_{\mathbf{H}}\} \leq\frac{C(\ln(\rho))^{N+1}}{\rho^N} \quad &\text{for } N=M,\\
&P\{\norm{\mathbf{w}}>c_M d_{\mathbf{H}}\} \leq\frac{C}{\rho^N} \quad &\text{for } N>M.
\end{align*}
Thus we have shown that augmented lattice reduction achieves the maximum receive diversity $N$ with the choice $t=\varepsilon a(\mathbf{H}_{\red})$.

\subsection{Simulation results} \label{simulations}

\begin{figure}[p] 
\begin{center}
\includegraphics[width=0.7\textwidth]{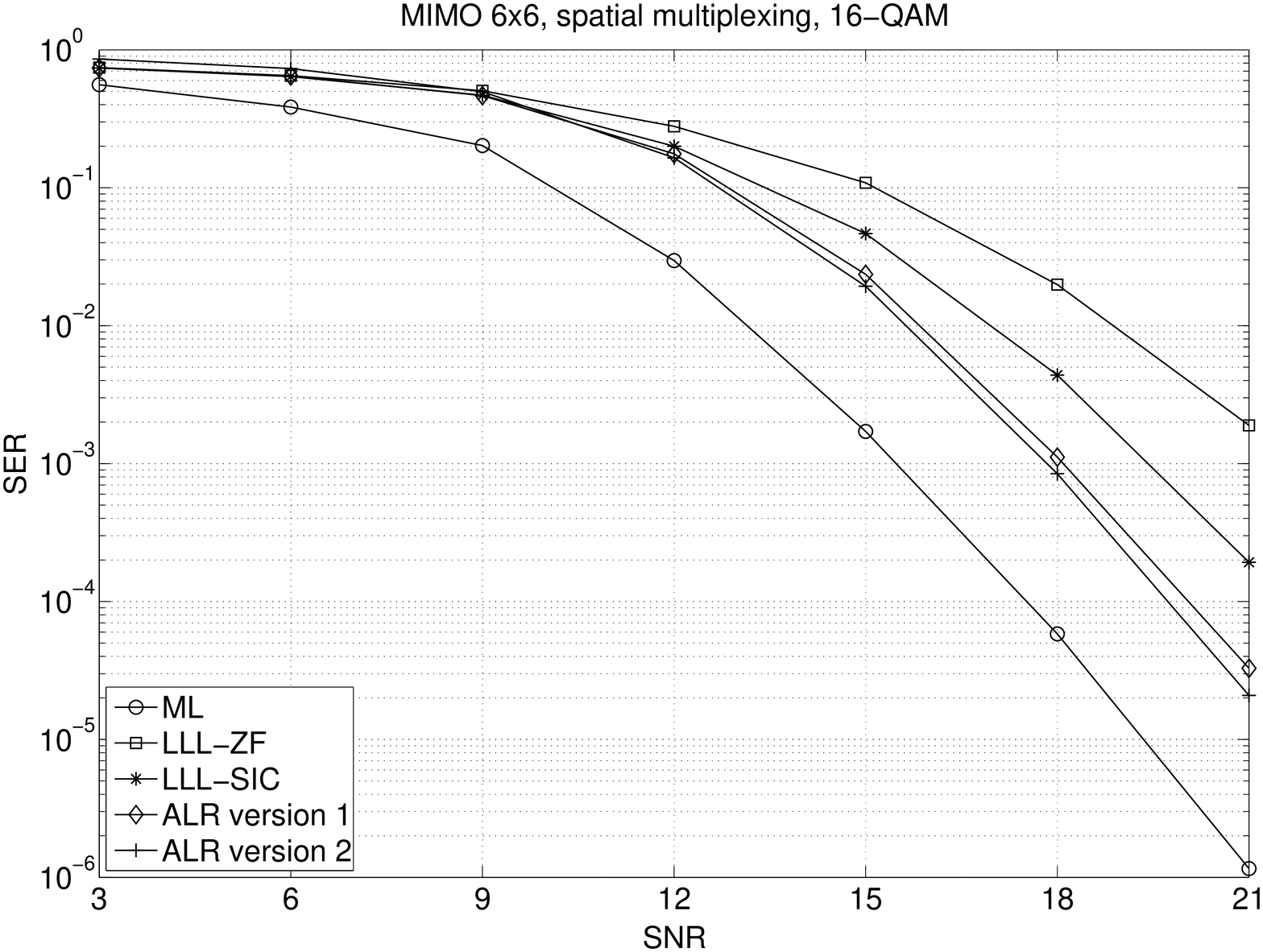}
\caption{Performance comparison of augmented lattice reduction with LLL-ZF and LLL-SIC detection for a $6 \times 6$ uncoded MIMO system using $16$-QAM. The LLL algorithm is performed using $\delta=\frac{3}{4}$.}
\label{6x6}
\end{center}
\end{figure}

Figure \ref{6x6} shows the performance of augmented lattice reduction for an uncoded $6 \times 6$ MIMO system using $16$-QAM constellations.\\
Two versions of augmented lattice reduction with different values of the parameter $\varepsilon$ are compared. Clearly it is preferable to choose $\varepsilon$ as big as possible in order to minimize the probability $P\{\norm{\mathbf{w}}> \varepsilon d_{\mathbf{H}}\}$. 
Version 1 corresponds to the choice $\varepsilon=\frac{1}{2\sqrt{2}\alpha^{m-\frac{1}{2}}}$, the highest value of $\varepsilon$ that verifies the hypothesis of Proposition \ref{receive_diversity}. At the SER of $2 \cdot 10^{-4}$, its performance is within $2.5\dB$ from ML decoding and gains $1.5\dB$ with respect to LLL-SIC decoding.\\
Version 2 corresponds to a value of $\varepsilon$ optimized by computer search (experimentally, this is around $2^{-\frac{m}{4}}$), whose performance is within $2.2\dB$ of ML decoding at the SER of $2 \cdot 10^{-4}$. From now on, we will always consider this optimized version.
For higher values of $\varepsilon$, we are not able to prove that the LLL algorithm finds the shortest lattice vector in $\mathcal{L}(\widetilde{\mathbf{H}})$. However, it is well-known that the LLL algorithm performs much better on average than the theoretical bounds predict.

\begin{figure}[p] 
\begin{center}
\includegraphics[width=0.7\textwidth]{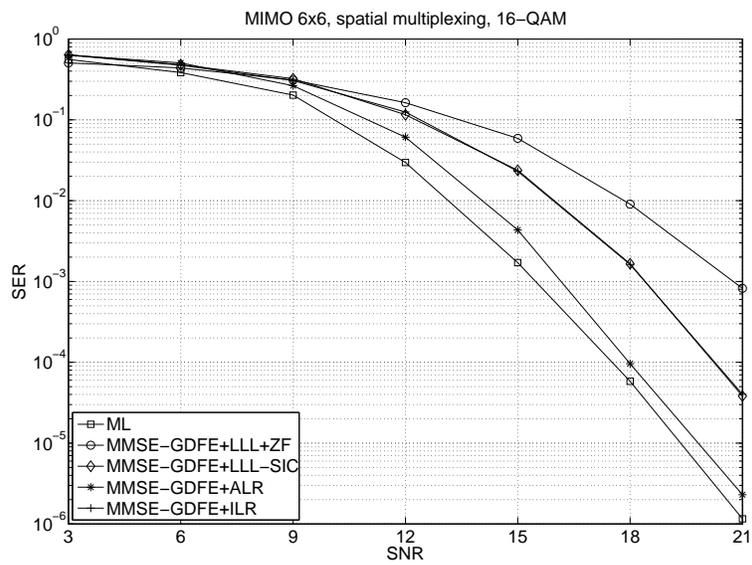}
\caption{Performance comparison of augmented lattice reduction with LLL-ZF, LLL-SIC and Improved Lattice Reduction with MMSE-GDFE preprocessing for a $6 \times 6$ uncoded MIMO system using $16$-QAM.}
\label{6x6withKP}
\end{center}
\end{figure}
\begin{figure}[p] 
\begin{center}
\includegraphics[width=0.7\textwidth]{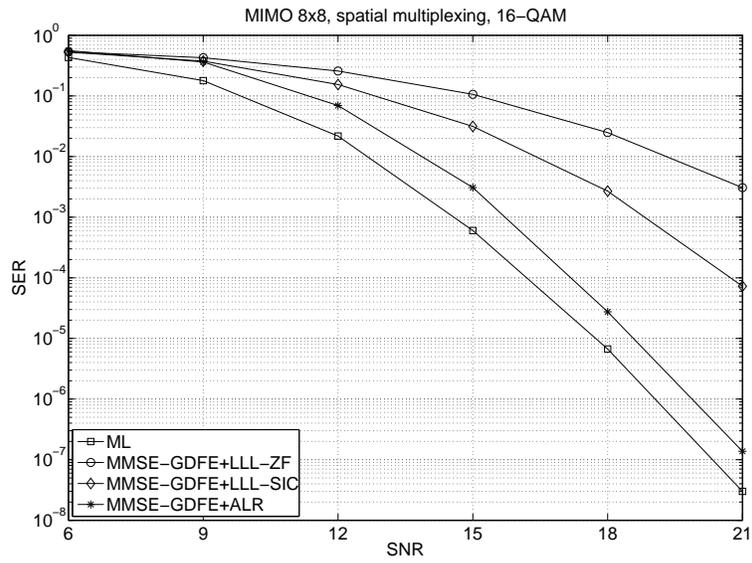}
\caption{Performance comparison of augmented lattice reduction with LLL-ZF and LLL-SIC detection with MMSE-GDFE preprocessing for a $8 \times 8$ uncoded MIMO system using $16$-QAM.}
\label{8x8_MMSE}
\end{center}
\end{figure}

In order to further reduce the distance from ML decoding, one can add MMSE-GDFE preprocessing, which yields a better conditioned channel matrix. Figure \ref{6x6withKP} shows the comparison of augmented lattice reduction with LLL-SIC detection, both using MMSE-GDFE preprocessing. At the SER of $10^{-4}$, augmented lattice reduction is within only $0.4\dB$ from ML performance and gains $2.3\dB$ with respect to LLL-SIC decoding.\\
The gain with respect to LLL-SIC decoding increases with the number of antennas: it is $3.5\dB$ for an $8\times8$ MIMO system, at the SER of $10^{-4}$. On the other side, augmented lattice reduction is still within $0.8\dB$ from ML performance (see Figure \ref{8x8_MMSE}).



\subsection{Comparison with Kim and Park's \vv{Improved Lattice Reduction}}
A lattice-reduction aided detection technique based on an augmented matrix similar to (\ref{Htilde}) (after MMSE-GDFE preprocessing) has been proposed in \cite{KP}. However, the philosophy behind the method of \cite{KP} is quite different: the parameter $t$ is chosen in such a way that the Lovasz condition on the last column of the augmented matrix is always verified. Specifically, considering the QR decomposition $\mathbf{H}_{\red}=\mathbf{QR}$ of the LLL-reduced matrix $\mathbf{H}_{\red}$, the condition $t>r_{m,m}$ is required. In general, this results in a much bigger value of the parameter $t$. Thus the transmitted message is detected from the last vector of the reduced augmented basis instead of the smallest basis vector.\\
On one side, this guarantees that the complexity increase is trivial because the only step required after reducing $\mathbf{H}$ is size reduction on the last column. On the other side, unlike our exponential gap technique, there is no guarantee that LLL reduction can find the required lattice vector. As a consequence, the performance of the decoder described in \cite{KP} is not as good, especially as the number of antennas increases; in fact it is about the same as LLL-SIC (see Figure \ref{6x6withKP}). 
The authors then propose to use a quantization error correction to improve the performance, which requires an additional computational cost, and is not needed in our case. \\


\section{Complexity} \label{complexity}
In this section we propose to estimate the additional complexity required by augmented lattice reduction with respect to LLL-ZF and LLL-SIC decoding. We are interested in the complexity order as a function of the number of transmit and receive antennas.

\subsection{Theoretical bounds} 
The complexity of LLL reduction of a gaussian channel matrix $\mathbf{H}$ has been studied in \cite{JSM}. As we have seen in Section \ref{preliminaries}, every instance of the LLL-ZF (respectively LLL-SIC) decoder consists of three main phases:
\newcounter{itemcounter}
\begin{enumerate}
\item A full \emph{Gram-Schmidt orthogonalization} is performed at the beginning of the LLL algorithm. This requires $O(nm^2)$ elementary operations \cite{GV}.
\item The \emph{main} "while" \emph{loop} of the LLL algorithm requires $O(m^2)$ elementary operations for each iteration. The number $K(\mathbf{H})$ of iterations of the LLL algorithm for a fixed realization $\mathbf{H}$ of the channel is bounded by \cite{JSM,DV} 
\begin{equation} \label{K}
K(\mathbf{H}) \leq m^2 \log_{\frac{1}{\sqrt{\delta}}} \left(\frac{A(\mathbf{H})}{a(\mathbf{H})}\right)+m,
\end{equation}
where $A(\mathbf{H})$ and $a(\mathbf{H})$ denote respectively the maximum and minimum norm of the Gram-Schmidt vectors of $\mathbf{H}$. For general $\mathbf{H}$,  $K(\mathbf{H})$ can be arbitrarily large. However, it was shown in \cite{JSM} that $\mathbb{E}(K(\mathbf{H}))\sim O\left(m^2\ln \left(\frac{m}{n-m+1}\right)\right)$. 
\item Finally, the \emph{ZF} and \emph{SIC receiver} entail respectively the multiplication by the pseudo-inverse of $\mathbf{H}_{\red}$ and its QR decomposition. Both have complexity order $O(nm^2)$ \cite{GV}.
\end{enumerate}
For fixed $\mathbf{H}$, we can use the estimate (\ref{K}) to obtain a bound of the number of iterations of the LLL reduction of $\widetilde{\mathbf{H}}$. The Gram-Schmidt orthogonalization of $\widetilde{\mathbf{H}}$ yields 
$$ \left(\begin{array}{cccc} \mathbf{h}_1^* & \cdots & \mathbf{h}_m^* & \mathbf{0}_{n \times 1} \\
 0 & \cdots & 0 & t \end{array}\right).$$
In fact, the last Gram-Schmidt vector is the projection of $\begin{pmatrix}-\mathbf{y} \\ t \end{pmatrix}$ on the  subspace  $$\left(\Span(\mathbf{h}_1^*,\ldots,\mathbf{h}_m^*)\right)^{\perp}\supseteq\left(\Span(\mathbf{e}_1,\ldots,\mathbf{e}_n)\right)^{\perp}=\Span(\mathbf{e}_{n+1}).$$ Therefore  
$$a(\widetilde{\mathbf{H}})\geq\min(t,a(\mathbf{H}))=\min(\varepsilon a(\mathbf{H}_{\red}),a(\mathbf{H})).$$ 
LLL reduction increases the minimum of the Gram-Schmidt vectors \cite{DV}, so $a(\mathbf{H}_{\red})\geq a(\mathbf{H})$, and $a(\widetilde{\mathbf{H}})\geq \varepsilon a(\mathbf{H})$. On the other side $t<a(\mathbf{H}_{\red}) \leq A(\mathbf{H}_{\red}) \leq A(\mathbf{H})$ and so $A(\widetilde{\mathbf{H}})=\max(t,A(\mathbf{H}))=A(\mathbf{H})$. Then
\begin{align*}
&K(\widetilde{\mathbf{H}}) \leq (m+1)^2 \log_{\frac{1}{\sqrt{\delta}}}\left(\frac{A(\widetilde{\mathbf{H}})}{a(\widetilde{\mathbf{H}})}\right)+m+1 \leq\\
& \leq \frac{(m+1)^2}{c} \ln\left(\frac{A(\mathbf{H})}{\varepsilon a(\mathbf{H})}\right)+m+1=\\
&= \frac{(m+1)^2}{c} \left(-\ln\varepsilon+\ln\left(\frac{A(\mathbf{H})}{a(\mathbf{H})}\right)\right)+m+1,
\end{align*}
where $c=\log{\frac{1}{\sqrt{\delta}}}$. Following \cite{JSM}, we can estimate the average $\mathbb{E}[K(\widetilde{\mathbf{H}})]$, recalling that $\frac{A(\mathbf{H})}{a(\mathbf{H})} \leq k(\mathbf{H})$, the condition number of $\mathbf{H}$, and that \cite{CD} 
$$\mathbb{E}[\ln k(\mathbf{H})] \leq \ln\left(\frac{m}{n-m+1}\right)+2.24.$$
We thus obtain
\begin{small}
\begin{align} \label{K_Lagarias}
&\mathbb{E}[K(\widetilde{\mathbf{H}})] \leq \frac{(m+1)^2}{c} \left(-\ln\varepsilon+\mathbb{E}[k(\mathbf{H})]\right)+m+1 \leq \notag \\
& \leq \frac{(m+1)^2}{c} \left(-\ln\varepsilon+\ln\left( \frac{m}{n-m+1}\right)+2.24\right)+m+1.  
\end{align}%
\end{small}%
For the choice $\varepsilon=\frac{1}{2\sqrt{2}\alpha^{m-\frac{1}{2}}}$, the complexity of the main loop of the LLL algorithm using the new method is at most of the order of $O(m^3)$. 

\subsection{Simulation results}
\begin{figure}[tbp] 
\begin{center}
\includegraphics[width=0.7\textwidth]{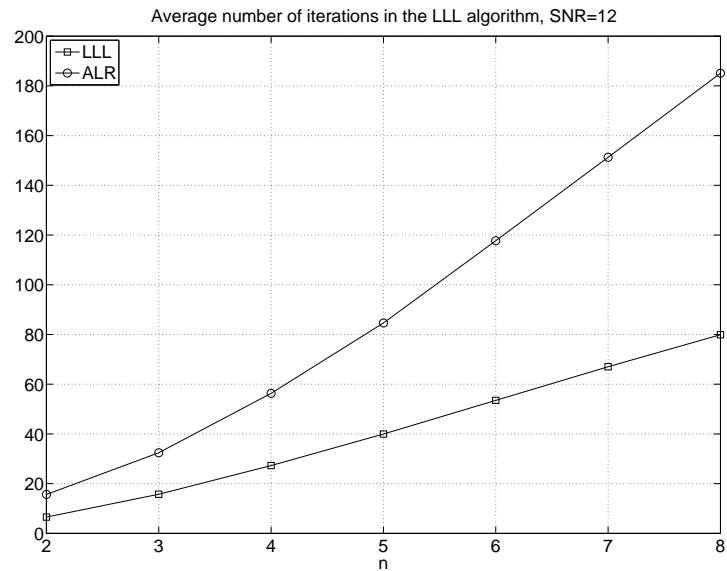}
\caption{Average number of steps of the LLL algorithm as a function of the number $n$ of transmit and receive antennas.}
\label{complexity_real}
\end{center}
\end{figure}

Our complexity simulations evidence the fact that the upper bounds (\ref{K}) and (\ref{K_Lagarias}) on the average number of iterations of the LLL algorithm for LLL-aided linear decoding and the augmented lattice reduction method are both quite pessimistic. The number of iterations for both methods appears in fact to be almost linear in practice, see Figure \ref{complexity_real}. \\
We have chosen $\delta=\frac{3}{4}$ in all the numerical simulations. 

\begin{figure}[tbp] 
\begin{center}
\includegraphics[width=0.7\textwidth]{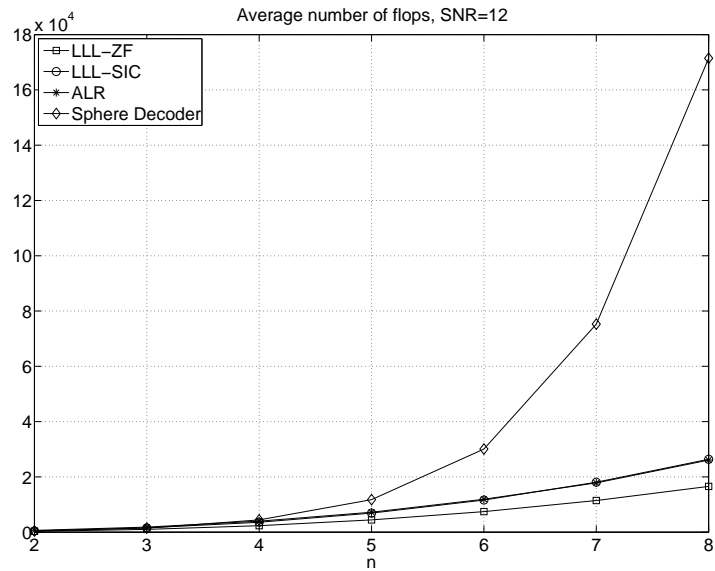}
\caption{Complexity comparison (in flops) of augmented lattice reduction with LLL-ZF, LLL-SIC and sphere decoding as a function of the number $n$ of transmit and receive antennas, at $\SNR=12$, using $16$-QAM constellations.}
\label{complexity_in_flops}
\end{center}
\end{figure}

While the number of iterations of LLL is indeed higher, approximately by a factor $2$, for the augmented lattice reduction (Figure \ref{complexity_real}), the total complexity expressed in flops\footnote{Here we define a \vv{flop} as any floating-point operation (addition, multiplication, division or square root).} is about the same for LLL-SIC and the augmented lattice method (see Figure \ref{complexity_in_flops}). The additional complexity of the LLL algorithm is balanced out by the complexity savings due to the fact that QR decomposition is not needed.


\subsection{Complex LLL reduction}
\begin{figure}[tbp] 
\begin{center}
\includegraphics[width=0.7\textwidth]{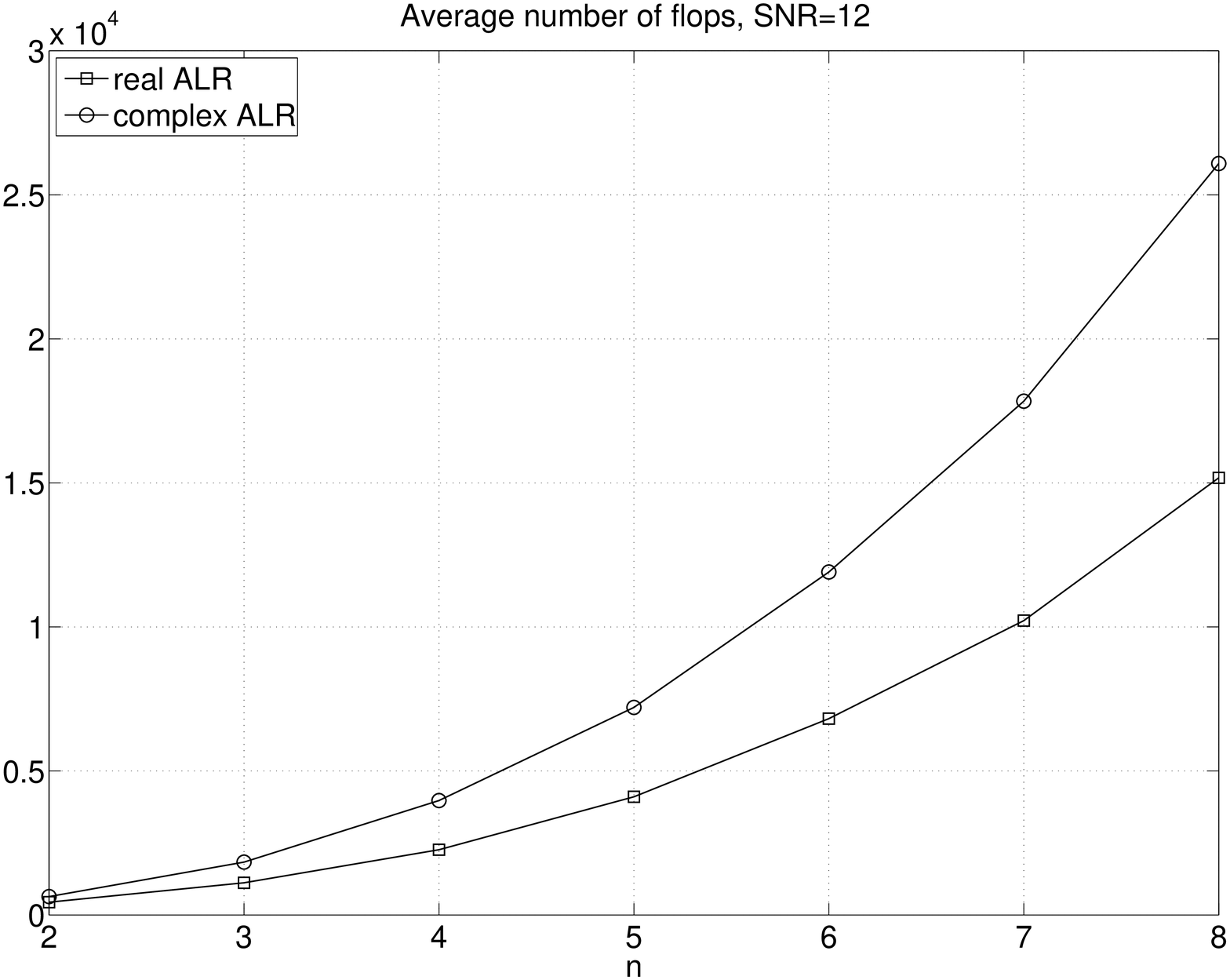}
\caption{Complexity comparison (in flops) of the real and complex version of augmented lattice reduction as a function of the number $n$ of transmit and receive antennas, at $\SNR=12$. Here we suppose that complex addition and complex multiplication require respectively $2$ and $6$ real flops.}
\label{complex_complexity}
\end{center}
\end{figure}

A generalization of the LLL algorithm to complex lattices has been studied in \cite{Nap} and applied to MIMO decoding in \cite{GLM}. It has been show experimentally in \cite{GLM} that the complex versions of LLL-ZF and LLL-SIC decoding have essentially the same performance of their real counterparts but with substantially reduced complexity.\\
A complex version of the augmented lattice reduction can be implemented by LLL-reducing the $(N+1) \times (M+1)$-dimensional matrix
$$\widetilde{\mathbf{H}}_{\cc}=\begin{pmatrix} \mathbf{H}_{\cc} & -\mathbf{y}_{\cc} \\ \mathbf{0}_{1\times N} & t \end{pmatrix},$$ and allows 
to save about $40\%$ of computational costs (see Figure \ref{complex_complexity}) without any change in performance.


\section{Conclusions}
In this paper, we introduced a new kind of lattice-reduction aided decoding  
which does not require a linear or decision-feedback receiver at the last stage. We proved that this method attains the maximum receive diversity order. Simulation results evidence that the new technique has a substantial performance gain with respect to the classical LLL-ZF and LLL-SIC decoders, while having approximately the same complexity order as LLL-SIC.


\small{

}

\begin{thebibliography}{40}
\itemsep=1mm
\parskip=0mm
\bibitem{Ba} L. Babai, \vv{On Lovasz' lattice reduction and the nearest lattice point problem}, \emph{Combinatorica}, vol. 6, n.1, pp 1--13 (1986)
\bibitem{CD} C. Chen, J.J. Dongarra, \vv{Condition numbers of Gaussian random matrices}, \emph{SIAM Journal on Matrix Analysis and Applications}, vol. 27, n.3 (2005), 603--620 
\bibitem{Co} H. Cohen, \vv{A course in computational algebraic number theory}, Graduate Texts in Mathematics, Springer, 2000
\bibitem{DEC2}  M. O. Damen, H. El Gamal, G. Caire, \vv{On maximum-likelihood detection and the search for the closest lattice point}, \emph{IEEE Trans. Inform. Theory}. vol. 49, 2389--2402, 2003 
\bibitem{DV} H. Daudé, B. Vallée, \vv{An upper bound on the average number of iterations of the LLL algorithm}, \emph{Theoretical Computer Science}, vol. 123, n.1 (1994), 95--115 
\bibitem{GLM} Y. H. Gan, C. Ling, W. H. Mow, \vv{Complex Lattice Reduction Algorithm for Low-Complexity MIMO Detection}, \emph{ IEEE Trans. Signal Process.}, vol 57 n.7 (2009)
\bibitem{GV} G.H. Golub, C.F. Van Loan, \vv{Matrix computations}, Johns Hopkins University Press, 1996
\bibitem{JE} J. Jaldén, P. Elia, \vv{DMT optimality of LR-aided linear decoders for a general class of channels, lattice designs, and system models}, submitted to \emph{IEEE Trans. Inform. Theory}
\bibitem{JSM} J. Jaldén, D. Seethaler, G. Matz, \vv{Worst- and average-case complexity of LLL lattice reduction in MIMO wireless systems}, \emph{IEEE International Conference on Acoustics, Speech and Signal Processing (ICASSP)}, (2008), 2685 -- 2688
\bibitem{Ka} R. Kannan, \vv{Minkowski's convex body theorem and integer programming}, \emph{Math. Oper. Res.} 12, 415--440 (1987)
\bibitem{KCM} K. Raj Kumar, G. Caire, A. L. Moustakas, \vv{Asymptotic performance of linear receivers in MIMO fading channels}, submitted.
\bibitem{KP} N. Kim, H. Park, \vv{Improved lattice reduction aided detections for MIMO systems}, \emph{Vehicular Technology Conference} 2006 
\bibitem{Li}  C. Ling, \vv{On the proximity factors of lattice reduction-aided decoding}, submitted.
\bibitem{LLL} A. K. Lenstra, J. H. W. Lenstra, L. Lovasz, \vv{Factoring polynomials with rational coefficients}, \emph{Math. Ann.}, vol. 261, pp. 515-534, 1982
\bibitem{LLS} J. C. Lagarias, H. W. Lenstra Jr., C. P. Schnorr, \vv{Korkin-Zolotarev bases and successive minima of a lattice and its reciprocal lattice}, \emph{Combinatorica}, vol. 10 n.4 (1990), 333--348 
\bibitem{Nap} H. Napias, \vv{A generalization of the LLL-algorithm over Euclidean rings or orders}, \emph{Journal de Th\'eorie des Nombres de Bordeaux} 8 (1996), 387-396
\bibitem{TMK} M. Taherzadeh, A. Mobasher, A. K. Khandani, \vv{LLL reduction achieves the receive diversity in MIMO decoding}, \emph{IEEE Trans. Inform. Theory}, vol 53 n. 12, 2007, pp 4801--4805
\bibitem{WF} C. Windpassinger, R. Fischer, \vv{Low-complexity near-maximum likelihood detection and precoding for MIMO systems using lattice reduction}, \emph{Proc IEEE Information Theory Workshop}, 2003, 345--348
\bibitem{YW} H. Yao, G. W. Wornell, \vv{Lattice-reduction-aided detectors for MIMO communication systems}, \emph{Proc. Global Telecommunications Conference} 2002, vol 1, 424--428
\end{thebibliography}
\end{document}